\documentstyle[mprocl]{article}

\def\be{\begin{eqnarray}}
\def\en{\end{eqnarray}}

\def\PRL{ Phys. Rev. Lett.}
\def\PRB{ Phys. Rev. {\bf  B}}

\begin{document}

\title{TWO-LOOP RENORMALIZATION
OF THE QUASIPARTICLE WEIGHT IN 
TWO-DIMENSIONAL  ELECTRON SYSTEMS}

\author{Jun-ichiro Kishine}

\address{Department of Theoretical Studies,
Institute for Molecular Science,
Okazaki 444-8585, Japan \\
E-mail: kishine@ims.ac.jp}

\author{Nobuo Furukawa}

\address{Department of Physics, Aoyama Gakuin University,
Setagaya, Tokyo 157-8572, Japan\\
E-mail: furukawa@phys.aoyama.ac.jp}

\author{Kenji Yonemitsu}

\address{Department of Theoretical Studies,
Institute for Molecular Science,
Okazaki 444-8585, Japan
\\
E-mail: kxy@ims.ac.jp}  

\maketitle\abstracts{We apply the renormalization-group (RG) approach
to two model systems  where 
the two-dimensional Fermi surface has portions which give
rise to the logarithmically singular two-loop self-energy process.}

\section{Introduction}
Non-Fermi-liquid behavior in the normal state of high $T_{\rm c}$ cuprates
has driven us to renewed interest in understanding the 
two-dimensional (2D) interacting electron systems in the framework of
the renormalization-group (RG) approach.\cite{Shankar94} 
In some cases, a 2D Fermi surface  has
portions which give rise to coupled infrared singularities in the particle-particle and
particle-hole  fluctuations.
Many attempts have so far been made at detecting
instabilities of the Landau-Fermi-Liquid  
by keeping track of the {\it one-loop}
RG flow of the corresponding susceptibilities. Among them are included 
the 2D systems with a  partially flat Fermi
surface,\cite{Zheleznyak97,Kishine99} tight-binding
dispersion,\cite{Zanchi96,Halboth99} and  
van-Hove singularity.\cite{Schulz87,Furukawa98b}
The breakdown 
of the Laudau-Fermi-liquid is also signaled by
the vanishing quasiparticle weight, $z$, which is treated on the {\it two-loop} level. 
In this paper, we apply the two-loop RG analysis
to two model systems 
where the 2D Fermi surface has portions which give
rise to the infrared logarithmic singularity (ILS) of the 
two-loop self-energy (2LSE).

\section{Fermi surface with partially flat regions}

As the first example,  
we consider a 2D Fermi surface shown in Fig.~1(a),
which  consists of   flat regions ($\alpha$, $\bar \alpha$,
$\beta$,   $\bar \beta$) with  length $2\Lambda$ and  round-arc 
regions.\cite{Zheleznyak97,Kishine99}
We linearlize the  one-particle dispersion with the bandwidth $E_0$
in the flat regions and parametrize the infrared cutoff as $\omega_l=E_{0}e^{-l}$ 
with a scaling parameter, $l$.
We assume $E_0\ll v_F\Lambda$ with 
$v_F$ being the corresponding Fermi velocity.
In this case, the RG equations are derived as in one dimension except that 
there appear infinite numbers of two-particle 
scattering vertices 
connecting the parallel flat regions [Fig.~1(b)].\cite{Zheleznyak97} 

\begin{figure}
\vskip 3cm
\caption{(a) The Fermi surface considered here.
(b) The two-particle scattering vertices connecting the flat regions $\alpha$ and
$\bar \alpha$, where $-\Lambda\leq k_{y1},k_{y2},k_{y3}\leq\Lambda$ and
$\sigma_{1},\cdots,\sigma_{4}$ are spin indices.
(c) The two-loop self-energy diagram.
The 
solid and broken lines represent the propagators for the  electrons in the
flat regions, $\alpha$ and $\bar \alpha$, respectively.}
\end{figure}

The RG equation for the quasiparticle weight at the position $k_y$ in the region $\alpha$, $z(k_y)$,
is written as\cite{Kishine99}
\begin{eqnarray}
{d \ln z(k_y)/ dl}= -\theta_l(k_y),
\end{eqnarray}
where $\theta_l(k_y)$  comes from the 2LSE diagram shown in Fig.~2(c)
and contains the scattering vertices which are also treated  on the two-loop level.
As an initial condition at   $l=0$,  we put the quasiparticle weight   equal 
to the non-interacting value, 
$z_{0}(k_y)=1,$
and the scattering strengths corresponding to the vertices  in Fig.~1(b)  equal 
to the  Hubbard repulsion, $U$.

In Fig.~2,  we show the  RG flow of $z(k_y)$
for $U=4.5\pi v_F$. We see that
the region  around which $z_{l}(k_y)$ is the most strongly suppressed moves
 from the center  ($k_y=0$) for $l<l_{\rm cross}\sim 3$ toward 
 the edges ($k_y=\pm \Lambda$) 
 for $l>l_{\rm cross}$ as the energy scale
 decreases and 
finally $z_{l}(k_y)$ approaches zero everywhere for $-\Lambda\leq k_y
\leq\Lambda$ in the low-energy limit.
The crossover behavior  originates
from  the breakdown of the cancellation between the particle-particle 
 and particle-hole  loops.\cite{Kishine99} 
 
\begin{figure}
\vskip 3cm
\caption{$k_y$ dependence of $z_{l}(k_y)$ at each RG step. }
\end{figure}

\section{Fermi surface which touches the umklapp surface}

As the second example,   we consider  a 2D  Fermi 
surface shown in Fig.~3(a), which touches the umklapp surface  at  4 points, $(\pm \pi/2, \pm \pi/2)$.\cite{HM93,Furukawa98a}
We divide the Fermi surface into patches and take account of
the electron one-particle and two-particle processes within   
 the patches
$\alpha=1,2,3,4$ 
around $(\pm \pi/2, \pm \pi/2)$ points [see Fig.~3(a)] 
which are connected through the normal
[$g_1$, $g_2$, $g_{1\rm r}$] and the umklapp [$g_3$, $g_{3\rm p}$, $g_{3\rm x}$]
processes,\cite{HM93,Furukawa98a} as indicated in Fig.~3(b).
We linearize the one-particle dispersion with the bandwidth $E_0$
 in the direction normal to
the Fermi surface at the center of the patch
and parametrize the infrared cutoff as $\omega_l=E_{0}e^{-l}$ 
with a scaling parameter, $l$.
In the energy scale, $\omega > \omega_{\rm T}$, where
$\omega_{\rm T}$ is   the infrared cutoff from the transverse dispersion,
the RG equations are derived as in one dimension except that there appear
additional inter-patch vertices, $g_{1\rm r}$ , $g_{3\rm p}$, and $g_{3\rm x}$.

\begin{figure}
\vskip 3cm
\caption{(a) The Fermi surface considered here.
(b) The two-particle scattering vertices entering the RG equations.
(c) The two-loop self-energy diagram.
Each leg of the vertices contains a patch index $\alpha=1,2,3,4$.}
\end{figure}

The 2LSE diagram, shown in Fig.~3(c), gives rise to the ILS for
 particular pairs of patch indices, $(\alpha,\beta)=(1,3)$ and (2,4). 
The RG equation for the quasiparticle weight at each patch, $z$,
is  
\begin{eqnarray}
d\ln z/dl=-(g_1^2+g_2^2-g_1g_2+g_3^2/2)/4.
\end{eqnarray}
The two-particle scattering vertices are also treated
on the two-loop level:
\begin{eqnarray}
dg_1/dl&=&-g_{1}^2-g_{1{\rm r}}^2-2g_{3{\rm x}}^2+2g_{3{\rm x}}g_{3{\rm p}}-g_{1}^3/2,\\
dg_2/dl&=&-(g_{1}^2+2g_{1{\rm r}}^2-g_{3}^2-2g_{3{\rm p}}^2)/2
-g_{1}^3/4+g_{3}^2(g_1-2g_2)/4,\\
dg_{1{\rm r}}/dl&=&-(g_1+g_2)g_{1{\rm r}},\\
dg_3/dl&=&-(g_{1}-2g_2)g_3-2g_{3{\rm x}}^2+2g_{3{\rm x}}g_{3{\rm p}}+g_{3{\rm p}}^2
-g_3(g_1-2g_2)^2/4-g_{3}^3/4,\\
dg_{3{\rm x}}/dl&=&-2g_{1}g_{3{\rm x}}+g_{1}g_{3{\rm p}}+g_{2}g_{3{\rm x}}
-g_{3}g_{3{\rm x}}+g_{3}g_{3{\rm p}},\\
dg_{3{\rm p}}/dl&=&(g_2+g_3)g_{3{\rm p}}.
\end{eqnarray}
The RG equations for $g_{1\rm r}$,  $g_{3\rm p}$,  and $g_{3\rm x}$ processes
do not contain two-loop contribution, on the same ground that
logarithmically singular contribution  of  the 2LSE is limited to
 particular pairs of patch indices. 
As an initial condition at  $l=0$,  we put the quasiparticle weight   equal 
to the non-interacting value, 
$z_{0}=1,$
and all the scattering strength corresponding to the
vertices in Fig.~3(b)  equal to the  Hubbard repulsion, $U$.

In Fig.~4(a), we show the two-loop RG flow of the scattering strengths
for $U=0.5\pi v_F$.
As in the case of the one-loop 
analysis,\cite{Furukawa98a}
all the  umklapp scatterings, $g_{3}$,  $g_{3\rm p}$, and $g_{3\rm x}$, 
are renormalized to a strong coupling sector  for $U>0$.
By examining the spin and charge susceptibilities,
Furukawa and Rice suggested that  the strong coupling fixed point 
corresponds to an insulating spin liquid state at the 4 patches
around $(\pm \pi/2, \pm \pi/2)$.

In Fig.~4(b) are shown the RG flows of the quasiparticle weight at the 4 patches, $z$,  in the following two cases:
{\bf (i)} $g_1$, $g_2$, and $g_3$ are present and other vertices are absent, and 
{\bf (ii)} all the vertices shown in Fig.~3(b) are present.
\begin{figure}
\vskip 3cm
\caption{(a) The two-loop RG flow of the scattering strengths
for $U=0.5\pi v_F$.
(b) The RG flows of $z$  in the cases where
{\bf (i)} $g_1$, $g_2$, and $g_3$ are present and other vertices are absent, and 
{\bf (ii)} all vertices shown in Fig.~3(b) are present.}
\end{figure}
The case {\bf (i)} is equivalent to the case of one dimension. 
We see that  the $g_{1\rm r}$, $g_{3\rm p}$, and $g_{3\rm x}$ processes,
which are peculiar to the 2D Fermi surface considered here, strongly suppress
the quasiparticle weight.
Consequently the quasiparticle weight at the 4 patches
approaches zero at a energy scale much higher than in the case {\bf (i)}. 
This result complements the view based on the one-loop analysis.\cite{Furukawa98a}

In both the first and second examples taken up here,
we have neglected effects of a Fermi surface curvature on the RG equations.
If we take  the finite curvature into account,
the RG flow of the quasiparticle weight  might stop 
around the energy scale $\omega_{\rm curv}$ corresponding to
the resolution  which detects  the curvature.
It remains matter for debate how to  incorporate the energy scale $\omega_{\rm curv}$
into the RG-based scheme in 2D. 

\section*{Acknowledgments}
J. K thanks J. B. Marston for helpful comments during the MBX conference.
N. F. thanks  T. M. Rice and M. Salmhofer for discussion. 
This work was supported by a Grant-in-Aid for Encouragement of Young
Scientists   from the Ministry of Education, Science, Sports and Culture,
Japan.

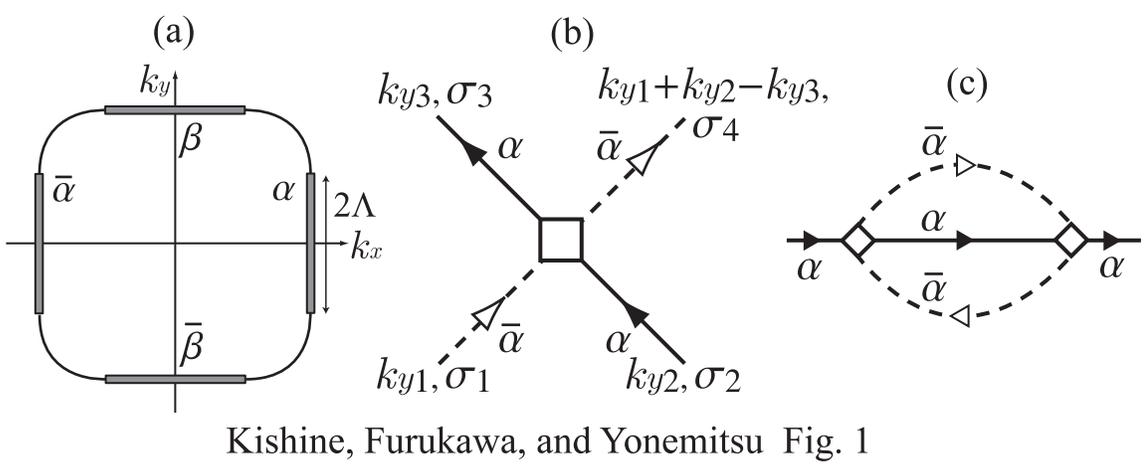
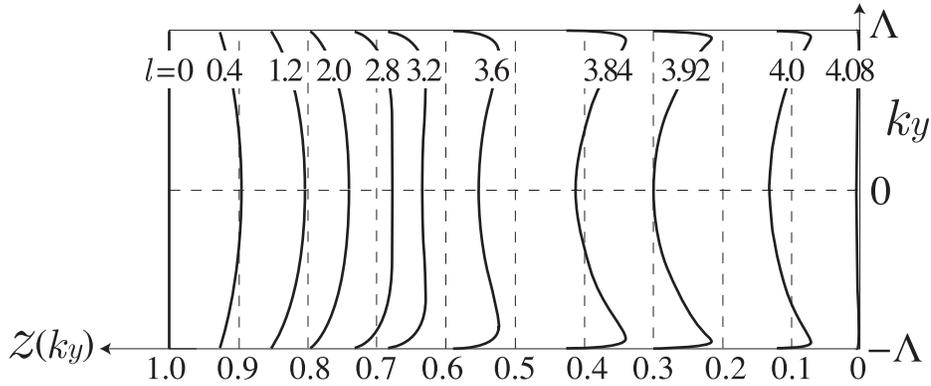
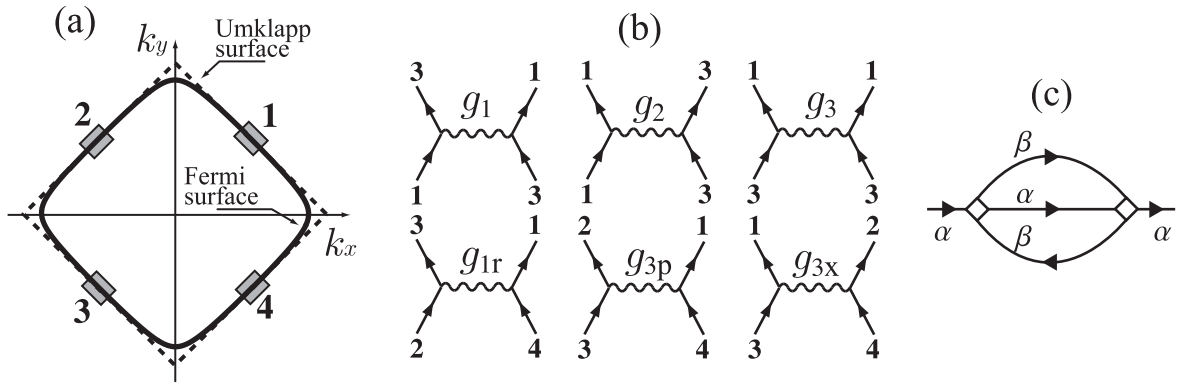
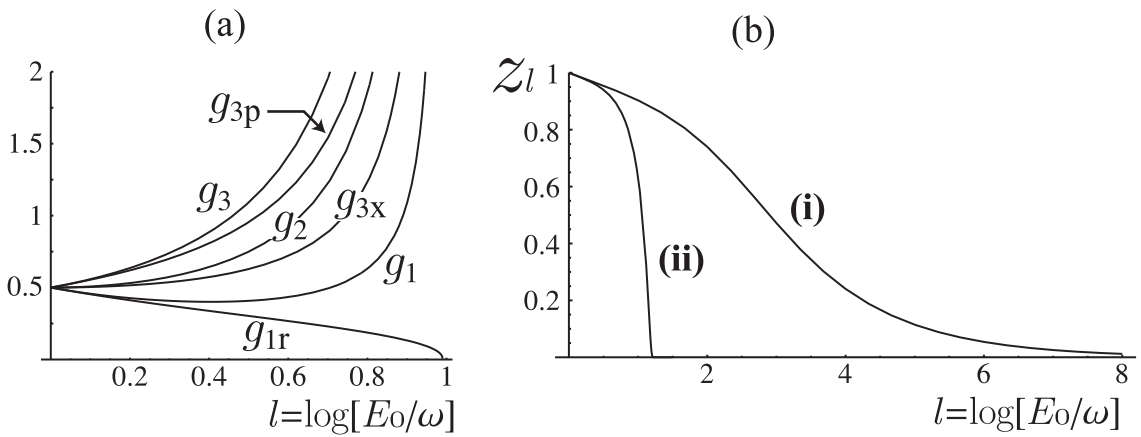
\end{document}